\documentstyle[aps,twoside]{revtex}
\oddsidemargin=\evensidemargin
\begin{document}
\twocolumn[\hsize\textwidth\columnwidth\hsize\csname@twocolumnfalse%
\endcsname
\draft
\title{Derivation of the Gasser-Leutwyler Lagrangian from QCD}
\author{Qing Wang$^{a,b}$, Yu-Ping Kuang$^{b,a}$,  Xue-Lei Wang$^{a,c}$,
Ming Xiao$^a$}
\address{$^a$Department of Physics, Tsinghua University, Beijing 100084, China
\footnote{Mailing address}
\\
$^b$China Center of Advanced Science and Technology (World Laboratory),
P.O.Box 8730, Beijing 100080, China\\
$^c$Department of Physics, Henan Normal University, Xinxiang 453002, China}
\date{TU-HEP-TH-99/105}
\bigskip

\maketitle

\begin{abstract}
The normal part of the Gasser-Leutwyler formulation of the chiral Lagrangian is
formally derived from the first principles of QCD without taking 
approximations. All the coefficients are expressed in terms of certain Green's
functions in QCD, which can be regarded as the fundamental QCD definitions of 
the normal part of the coefficients. The approximate values of the claculated
coefficients are also presented.
\end{abstract}


\vspace{0.2cm}
]

Chiral Lagrangian (CL)\cite{weinberg}\cite{GL}, especially the
three-flavor formulation given by Gasser and Leutwyler \cite{GL}, has been 
widely used in the study of low energy hadron physics. Studies on 
understanding the relation between the CL and the underlying theory of QCD 
will be very helpful for making the theory more predictive. There are papers 
studying approximate formulae for the CL \cite{Holdom,HW,Espriu} either
based on certain assumptions or considering the anomaly contributions 
with certain approximations from the beginning. Further improvements are 
certainly needed.
Actually, the study can be divided into two steps: (i) formally deriving the 
CL from QCD giving the fundamental QCD definitions of the coefficients, (ii) 
calculating the the values of the coefficients from the QCD definitions
with certain approximations. This paper is mainly devoted to step (i) in which 
we derive the fundamental QCD formulae from the normal part contributions of 
the theory without taking approximations, which is different from and 
complimentary to the anomaly part in Ref.\cite{Espriu}. We shall also present 
the values of our approximately calculated coefficients [step (ii)].

Let $A_{\mu}^i$ be the gluon field, ${\psi}_{\alpha}^{a\eta}$ and 
$\Psi_{\alpha}^{\bar{a}\eta}$ be, respectively, the light and heavy quark 
fields with color index ${\alpha}$, Lorentz spinor index $\eta$,
light flavor index $a$ and heavy flavor index $\bar{a}$. 
Following Gasser and Leutwyler, we start from the generating 
functional \cite{GL}
\begin{eqnarray}                                
&&Z[J]=\int{\cal D}\psi{\cal D}\bar{\psi}{\cal D}\Psi{\cal D}\bar{\Psi}
{\cal D}A_{\mu}
e^{i{\int}dx\{{\cal L}({\psi},\bar{\psi},\Psi,
\bar{\Psi},A_{\mu})+\bar{\psi}J\psi\}},\label{QCDZ}
\end{eqnarray}                                
where ${\cal L}$ is the QCD Lagrangian with the gauge-fixing term and
the Fadeev-Popov determinant. The local external sources $J_{\sigma\rho}$ 
can be decomposed as
\begin{eqnarray}                              
J(x)=-s(x) +ip(x)\gamma_5 +v\!\!\! /\;(x) +a\!\!\! /\;(x)\gamma_5 ,
\end{eqnarray}
in which the light quark masses have been absorbed into $s$. 
Since the contributions from the anomaly to the CL
has been studied in Ref.\cite{Espriu}, our aim is to study the complete 
normal part contributions, and thus we simply 
take the $\theta$-vacuum parameter $\theta=0$.

What we need to do is to {\it consistently} extract the pseudoscalar degree 
of freedom from the local composite operator $\bar \psi^{b\zeta}(x)
\psi^{a\eta}(x)$ and integrate out all the remaining degrees of freedom. For 
this purpose, we consider the scalar and pseudoscalar degrees of freedom of 
$\bar \psi^{b\zeta}(x)\psi^{a\eta}(x)$ and make the following decomposition
\begin{eqnarray}                                 
&&\bar{\psi}^{b\zeta}(x)(1)_{\zeta\eta}\psi^{a\eta}(x)=
(\Omega'\sigma\Omega'+\Omega^{\prime\dagger}
\sigma\Omega^{\prime\dagger})^{ab}(x)\nonumber\\
&&\bar{\psi}^{b\zeta}(x)(\gamma_5)_{\zeta\eta}\psi^{a\eta}(x)=
(\Omega'\sigma\Omega'-\Omega^{\prime\dagger}
\sigma\Omega^{\prime\dagger})^{ab}(x)
,\label{Udef0}
\end{eqnarray}
where the modular degree of freedom is described by an hermitian field 
$\sigma(x)$ [$\sigma^\dagger(x)=\sigma(x)$], and the phase degree of freedom
is described by a unitary field $\Omega'$ 
[$\Omega^{\prime\dagger}(x)\Omega^\prime(x)=1$]. 
As usual, we can define $U'(x)\equiv\Omega^{\prime 2}(x)$ which
contains a $U(1)$ factor such that det$U'(x)=e^{i\vartheta'(x)}$
($\vartheta'(x)$ is real). We can 
extract out the $U(1)$ factor and define a field $U(x)$ as $U'(x)\equiv 
e^{\frac{i}{N_f}\vartheta'(x)}U(x)$ with det$U(x)=1$. 
Then we can define a new field $\Omega$ and decompose $U$ into 
\begin{equation}                             
U(x)=\Omega^2(x)\label{Omegadef},
\end{equation}
as in the literature. This $U(x)$, as the desired representation of 
$SU(N_f)_R\times SU(N_f)_L$, will be the nonlinear realization of the meson 
field in the CL. It is straightforward to eliminate $\sigma$ from the two 
equations in (\ref{Udef0}) and get the following relations,
\begin{eqnarray}                            
e^{-i\frac{\vartheta'}{N_f}}
\Omega^{\dagger}{\rm tr}_l[P_R\psi\bar{\psi}]\Omega^{\dagger}
=e^{i\frac{\vartheta'}{N_f}}
\Omega{\rm tr}_l[P_L\psi\bar{\psi}]\Omega\label{relation},
\end{eqnarray}
\begin{equation}                           
e^{2i\vartheta'}={\rm det}[{\rm tr}_l[P_R\psi\bar{\psi}]]/
{\rm det}[{\rm tr}_l[P_L\psi\bar{\psi}]],\label{thetadef}
\end{equation}
in which all the fields are at the same space-time point $x$, and tr$_l$ is 
the trace for the spinor index. 
Eqs.(\ref{Udef0})-(\ref{thetadef}), especially (\ref{relation}), describe our 
idea of localization in the operator formalism. 

To realize this idea in the functional integration, we 
need a technique to {\it integrate in} this information to (\ref{QCDZ}). 
For this purpose, we start from the following functional identity for an 
operator ${\cal O}$ satisfying $\det{\cal O}=\det{\cal O}^\dagger$ \cite{WKWW1},
\begin{eqnarray}                             
&&\int{\cal D}U\;\delta(U^{\dagger}U-1)\delta({\rm det}U-1)
~{\cal F}[{\cal O}]~\delta(\Omega {\cal O}^{\dagger}\Omega-\Omega^{\dagger}
 {\cal O}\Omega^{\dagger})\nonumber\\
&&\hspace{0.4cm}={\rm const}\,\label{Uin},
\end{eqnarray}
in which $\int{\cal D}U\;\delta(U^{\dagger}U-1)\delta({\rm det}U-1)$ is an 
invariant integration measure and the function ${\cal F}[{\cal O}]$  is 
defined as
\begin{eqnarray}                        
\frac{1}{{\cal F}[{\cal O}]}\equiv{\rm det} {\cal O}~
\int{\cal D}\sigma\;
\delta({\cal O}^{\dagger}{\cal O}-\sigma^{\dagger}\sigma)
\delta(\sigma-\sigma^{\dagger})\,.\label{F}
\end{eqnarray}
With 
\begin{eqnarray}                         
{\cal O}(x)=e^{-i\frac{\vartheta'(x)}{N_f}}{\rm tr}_l[P_R\psi(x)\bar{\psi}(x)],
\label{O}
\end{eqnarray}
(\ref{Uin}) serves as the functional expression reflecting
(\ref{relation}).

Inserting (\ref{Uin}) and (\ref{O}) into (\ref{QCDZ}), we obtain
\begin{eqnarray}                              
&&Z[J]=\int{\cal D}\psi{\cal D}\bar{\psi}
{\cal D}\Psi{\cal D}\bar{\Psi}
{\cal D}A_{\mu}{\cal D}U~\delta(U^{\dagger}U-1)
\delta({\rm det}U-1)\nonumber\\
&&\hspace{0.4cm}\times\delta\bigg(e^{i\frac{\vartheta'}{N_f}}
\Omega{\rm tr}_l[\psi_L\bar{\psi}_R]\Omega-
e^{-i\frac{\vartheta'}{N_f}}
\Omega^{\dagger}{\rm tr}_l[\psi_R\bar{\psi}_L]\Omega^{\dagger}\bigg)
\nonumber\\
&&\hspace{0.4cm}\times\exp\big\{i\Gamma_I[\frac{\bar{\psi}\psi}{N_c}]
+i{\int}d^{4}x
\{{\cal L}(\psi,\bar{\psi},\Psi,\bar{\Psi},A_{\mu})
+\bar{\psi}J\psi\}\big\}
\nonumber\\
&&\hspace{0.4cm}=\int{\cal D}U\;\delta(U^{\dagger}U-1)
\delta({\rm det}U-1)\;e^{iS_{eff}[U,J]}\label{QCDZ'},
\end{eqnarray}
where 
\begin{eqnarray}                        
&&e^{-i\Gamma_I[\frac{1}{N_c}\bar{\psi}\psi]}
=\prod_x\frac{1}{{\cal F}[{\cal O}(x)]}\nonumber\\
&&\hspace{0.3cm}=\prod_x\big\{
\big[{\rm det}\{\frac{1}{N_c}{\rm tr}_{lc}[\psi_R\bar{\psi}_L]\}
{\rm det}\{\frac{1}{N_c}{\rm tr}_{lc}[\psi_L\bar{\psi}_R]\}
\big]^{\frac{1}{2}}
\int{\cal D}\sigma\nonumber\\
&&\hspace{0.5cm}\times\delta\big(
\frac{1}{N_c^2}{\rm tr}_{lc}(\psi_R\bar{\psi}_L)
{\rm tr}_{lc}(\psi_L\bar{\psi}_R)
-\sigma^{\dagger}\sigma\big)\delta(\sigma-\sigma^{\dagger})\big\},
\label{GammaTK}
\end{eqnarray}
and we have introduced the effective action
\begin{eqnarray}                             
&&e^{iS_{eff}[U,J]}=\int{\cal D}\psi{\cal D}\bar{\psi}{\cal D}
\Psi{\cal D}\bar{\Psi}{\cal D}A_{\mu}
\delta(e^{i\frac{\vartheta'}{N_f}}
\Omega{\rm tr}_l[\psi_L\bar{\psi}_R]\Omega\nonumber\\
&&\hspace{0.4cm}-e^{-i\frac{\vartheta'}{N_f}}
\Omega^{\dagger}{\rm tr}_l[\psi_R\bar{\psi}_L]\Omega^{\dagger})
\exp\bigg\{i\Gamma_I[\frac{\bar{\psi}\psi}{N_c}]\nonumber\\
&&\hspace{0.4cm}+i{\int}d^{4}x
\{{\cal L}(\psi,\overline{\psi},\Psi,\overline{\Psi},A_{\mu})
+\overline{\psi}J\psi\}\bigg\}
\label{Seffdef}
\end{eqnarray}
to formally express the integration over the quark and gluon fields.

To work out the $U(x)$-dependence of $S_{eff}$, we make the following
chiral rotation
\begin{eqnarray}                                     
J_{\Omega}(x)&=&[\Omega(x)P_R+\Omega^{\dagger}(x)P_L]
[J(x)+i\partial\!\!\!\! /\;]\nonumber\\
&&\times[\Omega(x)P_R+\Omega^{\dagger}(x)P_L],\nonumber\\
\psi_\Omega(x)&=&[\Omega^\dagger(x)P_R+\Omega(x)P_L]\psi(x),\nonumber\\
\bar{\psi}_\Omega(x)&=&\bar{\psi}(x)[\Omega^\dagger(x)P_R
+\Omega(x)P_L].\label{rotation}
\end{eqnarray}
Here we have denoted the rotated quantities by a subscript $\Omega$. The 
fields $U,A_{\mu},\Psi$ and $\bar{\Psi}$ are unchanged under the rotation.
It is easiy to see from (\ref{thetadef}) that $\vartheta'_\Omega=\vartheta'$.
In terms of the rotated fields, the $\delta$-function in
(\ref{Seffdef}) is $\delta\bigg(e^{i\frac{\vartheta'}{N_f}}{\rm tr}_l
[\psi_{\Omega,L}\bar{\psi}_{\Omega,R}]-e^{-i\frac{\vartheta'}{N_f}}{\rm tr}_l
[\psi_{\Omega,R}\bar{\psi}_{\Omega,L}]\bigg)$, so that the explicit
$U$-dependence of the theory is from the source term and the
$\delta$-function in (\ref{Seffdef}).
The remaining part in (\ref{Seffdef}) is invariant under rotation
(\ref{rotation}) since the Lagrangian ${\cal L}$, together with the
source-term, is chirally invariant, and from (\ref{rotation}) we can see that
$\Gamma\big[\frac{\bar\psi\psi}{N_c}\big]
=\Gamma\big[\frac{\bar \psi_\Omega\psi_\Omega}{N_c}\big]$.
Since the integration measure ${\cal D}\psi{\cal D}\bar \psi$ is not
invariant under the chiral rotation, it will cause certain anomaly
terms in $S_{eff}$ i.e.
$\int{\cal D}\psi{\cal D}\overline{\psi}{\cal D}\Psi{\cal D}\overline{\Psi}
{\cal D}A_{\mu}~
=\int{\cal D}\psi_{\Omega}{\cal D}\overline{\psi}_{\Omega}
{\cal D}\Psi{\cal D}\overline{\Psi}{\cal D}A_{\mu}~
e^{[\mbox{anomaly terms}]}$.
Thus, making a chiral rotation of $\psi$ and $\bar \psi$, we have
\begin{eqnarray}                             
&&e^{iS_{\rm eff}[1,J_\Omega]}
=\int{\cal D}\psi{\cal D}\bar{\psi}{\cal D}\Psi{\cal D}\bar{\Psi}
{\cal D}A_{\mu}~
\delta\bigg(e^{i\frac{\vartheta'}{N_f}}
{\rm tr}_l[\psi_L\bar{\psi}_R]\nonumber\\
&&\hspace{0.4cm}-e^{-i\frac{\vartheta'}{N_f}}
{\rm tr}_l[\psi_R\bar{\psi}_L]\bigg)
\exp \big\{i\Gamma_I[\frac{\bar{\psi}\psi}{N_c}]\nonumber\\
&&\hspace{0.4cm}+i{\int}d^{4}x
[{\cal L}(\psi,\bar{\psi},\Psi,\bar{\Psi},A_{\mu})
+\bar{\psi}J_{\Omega}\psi]+\mbox{anomaly terms}\big\}\nonumber\\
&&\hspace{0.2cm}=\int{\cal D}\psi{\cal D}\overline{\psi}{\cal D}\Psi{\cal D}
\bar{\Psi}{\cal D}A_{\mu}~\delta\bigg(\bar{\psi}^a(-i\sin\frac{\vartheta'}{N_f}
+\gamma_5\cos\frac{\vartheta'}{N_f})\psi^b\bigg)
\nonumber\\
&&\hspace{0.4cm}\times\exp \big\{i\Gamma_I[\frac{\bar{\psi}\psi}{N_c}]
+i{\int}d^{4}x
[{\cal L}(\psi,\overline{\psi},\Psi,\bar{\Psi},A_{\mu})+\bar \psi
(v\!\!\! /_{\Omega}\nonumber\\
&&\hspace{0.4cm}+a\!\!\! /_{\Omega}\gamma_5
-s_{\Omega}-p_{\Omega}\tan\frac{\vartheta'}{N_f})\psi]
+\mbox{anomaly terms}\big\}.\label{Sefffinal}
\end{eqnarray}
In (\ref{Sefffinal}) and all the later formulae, we simply use the symbol
$\psi$ ($\bar \psi$) as the short notations of $\psi_\Omega$ 
($\bar \psi_\Omega$) for shortening the formulae. This does not make
confusion since they are integration variables). We see that, in 
(\ref{Sefffinal}), the $U$-dependence of $S_{eff}$ comes only from the 
rotated sources. So that the rotation simplifies the $U$-dependence of 
$S_{eff}$. Eq.(\ref{Sefffinal}) shows that $S_{eff}[1,J_\Omega]$ is the QCD 
generating functional for the rotated sources with the 
$-i\bar{\psi}^a\psi^b\sin\frac{\vartheta'}{N_f}$
$+\bar{\psi}^a\gamma_5\psi^b\cos\frac{\vartheta'}{N_f}$
degree of freedom frozen. After making a further $U_A(1)$ rotation of the 
$\psi$ and $\bar{\psi}$, $\frac{\vartheta'}{N_f}$ can be rotated away and the 
frozen degree of freedom just becomes the pesudoscalar degree of freedom 
$\bar{\psi}^a\gamma_5\psi^b$ as it should be
since this degree of freedom is already included in the $U$-field. The 
automatic occurance of this frozen degree of freedom implies that our way of 
extracting the $U$-field degree of freedom is really {\it consistent}.

We first consider the $p^2$-order terms.  
Expanding (\ref{Sefffinal}) up to the order of $p^2$, and taking
account of translational invariance, parity and flavor conservations, we obtain
\begin{eqnarray}                      
&&S_{eff}[1,J_{\Omega}]
\bigg|_{O(p^2)}
=F_0^2\int dx~{\rm tr}[a_{\Omega}^2+B_0s_{\Omega}]
\nonumber\\ 
&&=\frac{F_0^2}{4}\int dx~{\rm tr}
\bigg[[\nabla^{\mu}U^{\dagger}][\nabla_{\mu}U]+[U\chi^\dagger+U^\dagger\chi]
\bigg],\label{CLp2}
\end{eqnarray}
where $\nabla_\mu$ is the covariant derivative related to the external
sources defined in Ref.\cite{GL}, $\chi\equiv 2B_0(s+ip)$, and the 
coefficients $F^2_0$ and $B_0$ are defined in terms of the following QCD 
Green's functions as 
\begin{eqnarray}                            
F_0^2B_0\equiv -\frac{1}{N_f}
\bigg\langle\bar{\psi}\psi\bigg\rangle,
\label{F0B0def}
\end{eqnarray}
and
\begin{eqnarray}                                
&&F_0^2
\equiv\frac{i}{8(N_f^2-1)}\int dx\bigg[
\bigg\langle\overline{\psi}^a(0)\gamma^{\mu}\gamma_5\psi^b(0)
\bar{\psi}^b(x)\gamma_{\mu}\gamma_5\psi^a(x)\bigg\rangle_C
\label{F0def}\nonumber
\end{eqnarray}
\vspace{-0.5cm}
\begin{eqnarray}
&&-\frac{1}{N_f}
\bigg\langle[\overline{\psi}^a(0)\gamma^{\mu}\gamma_5\psi^a(0)]
[\overline{\psi}^b(x)\gamma_{\mu}\gamma_5\psi^b(x)]\bigg\rangle_C\bigg]\,.
\end{eqnarray}
In (\ref{F0B0def}) and (\ref{F0def}), the vacuum expectation values are
defined as, for an operator $O$,
\begin{eqnarray}                                
&&\displaystyle
\bigg\langle O\bigg\rangle\equiv
\frac{\int{\cal D}\mu~O}{\int{\cal D}\mu},\nonumber\\
&&{\cal D}\mu \equiv {\cal D}\psi{\cal D}
\bar{\psi}{\cal D}\Psi{\cal D}\bar{\Psi}{\cal D}A_{\mu}
\delta\bigg(\bar{\psi}^a\big(-i\sin\frac{\vartheta'}{N_f}
+\gamma_5\cos\frac{\vartheta'}{N_f}\big)\psi^b\bigg)\nonumber
\end{eqnarray}
\vspace{-0.5cm}
\begin{eqnarray}
&&\times e^{i\Gamma_I[\frac{\bar{\psi}\psi}{N_c}]+
i\int dx{\cal L}({\psi},\bar{\psi},\Psi,\bar{\Psi},A_{\mu})}\,,\hspace{1cm}
\label{average}
\end{eqnarray}
and $\langle\cdots\rangle_C$ denotes the connected part of 
$\langle\cdots\rangle$.
The integrand in (\ref{CLp2}) is just the $p^2$-order terms in the 
Gasser-Leutwyler Lagrangian \cite{GL}. 

The $p^4$-order terms can be worked out along the same line. With the
help of the $p^2$-order equation of motion, we obtain the normal part 
contributions (ignoring anomaly contributions) to the $p^4$-order terms,
\begin{eqnarray}                        
&&S_{eff}[1,J_{\Omega}]\bigg|_{O(p^4)}=\int dx~{\rm tr}\bigg[
-{\cal K}_1[d_{\mu}a_{\Omega}^{\mu}]^2\nonumber\\
&&-{\cal K}_2(d^{\mu}a_{\Omega}^{\nu}-d^{\nu}a_{\Omega}^{\mu})
(d_{\mu}a_{\Omega,\nu}-d_{\nu}a_{\Omega,\mu})
+{\cal K}_3[a_{\Omega}^2]^2\nonumber\\
&&+{\cal K}_4a_{\Omega}^{\mu}a_{\Omega}^{\nu}a_{\Omega,\mu}a_{\Omega,\nu}
+{\cal K}_5a_\Omega^2{\rm tr}[a_{\Omega}^2]
+{\cal K}_6a_{\Omega}^{\mu}a_{\Omega}^{\nu}
{\rm tr}[a_{\Omega,\mu}a_{\Omega,\nu}]\nonumber\\
&&+{\cal K}_7s_{\Omega}^2
+{\cal K}_8s_{\Omega}{\rm tr}[s_{\Omega}]+{\cal K}_9p_{\Omega}^2
+{\cal K}_{10}p_{\Omega}{\rm tr}[p_{\Omega}]
+{\cal K}_{11}s_{\Omega}a_{\Omega}^2
\nonumber\\
&&+{\cal K}_{12}s_{\Omega}{\rm tr}
[a_{\Omega}^2]
-{\cal K}_{13}V_{\Omega}^{\mu\nu}V_{\Omega,\mu\nu}
+i{\cal K}_{14}V_{\Omega}^{\mu\nu}a_{\Omega,\mu}a_{\Omega,\nu}\nonumber\\
&&+{\cal K}_{15}p_\Omega d^\mu a_{\Omega,\mu}\bigg]
\label{p4}\nonumber\\
&&=\int dx\bigg[L_1[{\rm tr}(\nabla^{\mu}U^{\dagger}\nabla_{\mu}U)]^2
\nonumber\\
&&+L_2{\rm tr}[\nabla_{\mu}U^{\dagger}\nabla_{\nu}U]
{\rm tr}[\nabla^{\mu}U^{\dagger}\nabla^{\nu}U]
\nonumber\\
&&
+L_3{\rm tr}[(\nabla^{\mu}U^{\dagger}\nabla_{\mu}U)^2]
+L_4{\rm tr}[\nabla^{\mu}U^{\dagger}\nabla_{\mu}U]
{\rm tr}[\chi^{\dagger}U+U^\dagger\chi]\nonumber\\
&&
+L_5{\rm tr}[\nabla^{\mu}U^{\dagger}\nabla_{\mu}U(\chi^{\dagger}U
+U^\dagger\chi)]
+L_6[{\rm tr}(\chi^{\dagger}U+U^\dagger\chi)]^2\nonumber\\
&&+L_7[{\rm tr}(\chi^{\dagger}U-U^\dagger\chi)]^2
+L_8{\rm tr}[\chi^{\dagger}U\chi^{\dagger}U+\chi U^\dagger\chi U^\dagger]
\nonumber\\
&&
-iL_9{\rm tr}[F_{\mu\nu}^R\nabla^{\mu}U\nabla^{\nu}U^{\dagger}
+F_{\mu\nu}^L\nabla^{\mu}U^\dagger\nabla^{\nu}U]\nonumber\\
&&+L_{10}{\rm tr}[U^{\dagger}F_{\mu\nu}^RUF^{L,\mu\nu}]
+H_1{\rm tr}[F_{\mu\nu}^RF^{R,\mu\nu}+F_{\mu\nu}^LF^{L,\mu\nu}]
\nonumber\\
&&
+H_2{\rm tr}[\chi^{\dagger}\chi]\bigg],\label{p4GL}
\end{eqnarray}
where $d^\mu a_\Omega^\nu\equiv \partial^\mu a_\Omega^\nu-iv_\Omega^\mu
a_\Omega^\nu+ia_\Omega^\mu v_\Omega^\nu,~~V_\Omega^{\mu\nu}\equiv
\partial^\mu v_\Omega^\nu-\partial^\nu v_\Omega^\mu-iv_\Omega^\mu
v_\Omega^\nu+iv_\Omega^\nu v_\Omega^\mu$,  
and
\begin{eqnarray}                          
&&L_1=\frac{{\cal K}_4+2{\cal K}_5+2{\cal K}_{13}-{\cal K}_{14}}{32},
\nonumber\\
&&L_2=\frac{{\cal K}_4+{\cal K}_6+2{\cal K}_{13}-{\cal K}_{14}}{16},
\nonumber\\
&&L_3=\frac{{\cal K}_3-2{\cal K}_4-6{\cal K}_{13}+3{\cal K}_{14}}{16},~
L_4=\frac{{\cal K}_{12}}{16B_0},~
\nonumber\\
&&L_5=\frac{{\cal K}_{11}}{16B_0},~
L_6=\frac{{\cal K}_8}{16B_0^2},~
L_7=-\frac{B^2_0{\cal K}_1+N_f{\cal K}_{10}+B_0{\cal K}_{15}}{16B^2_0N_f},
\nonumber\\
&&
L_8=\frac{B^2_0{\cal K}_1+{\cal K}_7-{\cal K}_9+B_0{\cal K}_{15}}{16B^2_0},
L_9= \frac{4{\cal K}_{13}-{\cal K}_{14}}{8},~
\nonumber\\
&&
L_{10}=-\frac{{\cal K}_2+{\cal K}_{13}}{2},~
H_1=\frac{{\cal K}_2-{\cal K}_{13}}{4},~
\nonumber\\
&&
H_2=\frac{-B^2_0{\cal K}_1+{\cal K}_7+{\cal K}_9-B_0{\cal K}_{15}}{8B^2_0}.
\label{p4C}
\end{eqnarray}
The coefficients $~{\cal K}_1,\cdots {\cal K}_{15}$ are defined by the 
following integrations of the QCD Green's functions
\begin{eqnarray}                            
&&\frac{i}{4}\int dx\; x^{\mu'}x^{\nu'}
\bigg\langle[\bar{\psi}^a(0)\gamma^{\mu}\gamma_5\psi^b(0)]
[\bar{\psi}^c(x)\gamma^{\nu}\gamma_5\psi^d(x)]\bigg\rangle_C
\nonumber\\
&&=[(\frac{{\cal K}_1}{2}-{\cal K}_2)(g^{\mu\mu'}g^{\nu\nu'}+g^{\mu\nu'}
g^{\nu\mu'})
+2{\cal K}_2g^{\mu\nu}g^{\mu'\nu'}]\delta^{ad}\delta^{bc}
\nonumber\\
&&+\cdots\nonumber\\
&&-\frac{i}{24}\int dxdydz
\bigg\langle[\bar{\psi}^{a}(0)\gamma^{\mu}\gamma_5\psi^{b}(0)]
[\bar{\psi}^{c}(x)\gamma^{\nu}\gamma_5\psi^{d}(x)]\nonumber\\
&&\times[\bar{\psi}^{e}(y)\gamma^{\lambda}\gamma_5\psi^{f}(y)]
[\bar{\psi}^{g}(z)\gamma^{\kappa}\gamma_5\psi^{h}(z)]\bigg\rangle_C
\nonumber\\
&&=\frac{1}{6}\bigg\{\delta^{ad}\bigg[
\delta^{cf}\delta^{eh}\delta^{gb}[
\frac{1}{2}(g^{\mu\nu}g^{\lambda\kappa}
+g^{\mu\kappa}g^{\nu\lambda}){\cal K}_3
+g^{\mu\lambda}g^{\nu\kappa}{\cal K}_4]
\nonumber\\
&&+\delta^{ch}\delta^{gf}\delta^{eb}[
\frac{1}{2}(g^{\mu\nu}g^{\lambda\kappa}
+g^{\mu\lambda}g^{\nu\kappa}){\cal K}_3
+g^{\mu\kappa}g^{\nu\lambda}{\cal K}_4]\bigg]
\nonumber\\
&&+\delta^{af}\bigg[
\delta^{ed}\delta^{ch}\delta^{gb}[
\frac{1}{2}(g^{\mu\lambda}g^{\nu\kappa}
+g^{\mu\kappa}g^{\nu\lambda}){\cal K}_3
+g^{\mu\nu}g^{\lambda\kappa}{\cal K}_4]\nonumber\\
&&+\delta^{eh}\delta^{gd}\delta^{cb}[
\frac{1}{2}(g^{\mu\lambda}g^{\nu\kappa}
+g^{\mu\nu}g^{\lambda\kappa}){\cal K}_3
+g^{\mu\kappa}g^{\nu\lambda}{\cal K}_4]\bigg]\nonumber\\
&&+\delta^{ah}\bigg[\delta^{gd}\delta^{cf}\delta^{eb}[
\frac{1}{2}(g^{\mu\kappa}g^{\nu\lambda}
+g^{\mu\lambda}g^{\nu\kappa}){\cal K}_3
+g^{\mu\nu}g^{\lambda\kappa}{\cal K}_4]\nonumber\\
&&+\delta^{gf}\delta^{ed}\delta^{cb}[
\frac{1}{2}(g^{\mu\kappa}g^{\nu\lambda}
+g^{\mu\nu}g^{\lambda\kappa}){\cal K}_3
+g^{\mu\lambda}g^{\nu\kappa}{\cal K}_4]\bigg]\nonumber\\
&&+\delta^{ad}\delta^{cb}\delta^{eh}\delta^{gf}
[g^{\mu\nu}g^{\lambda\kappa}2{\cal K}_5
+(g^{\mu\lambda}g^{\nu\kappa}+g^{\mu\kappa}g^{\nu\lambda}){\cal K}_6]
\nonumber\\
&&+\delta^{af}\delta^{eb}\delta^{ch}\delta^{gd}
[g^{\mu\lambda}g^{\nu\kappa}2{\cal K}_5
+(g^{\mu\nu}g^{\lambda\kappa}+g^{\mu\kappa}g^{\nu\lambda}){\cal K}_6]
\nonumber\\
&&+\delta^{ah}\delta^{gb}\delta^{cf}\delta^{ed}
[g^{\mu\kappa}g^{\nu\lambda}2{\cal K}_5
+(g^{\mu\nu}g^{\lambda\kappa}+g^{\mu\lambda}g^{\nu\kappa}){\cal K}_6]\bigg\}
+\cdots\nonumber\\
&&\frac{i}{2}\int dx\bigg\langle[\bar{\psi}^a(0)\psi^b(0)]
[\bar{\psi}^c(x)\psi^d(x)]\bigg\rangle_C
\nonumber\\
&&={\cal K}_7\delta^{ad}\delta^{bc}
+{\cal K}_8\delta^{ab}\delta^{cd}\nonumber\\
&&\frac{i}{2}\int dx\bigg\langle[\bar{\psi}^a(0)\psi^b(0)]
\tan\frac{\vartheta'(0)}{N_f}[\bar{\psi}^c(x)\psi^d(x)]
\tan\frac{\vartheta'(x)}{N_f}
\bigg\rangle_C
\nonumber\\
&&={\cal K}_9\delta^{ad}\delta^{bc}+{\cal K}_{10}\delta^{ab}\delta^{cd}
\nonumber\\
&&\frac{1}{8}\int dxdy\bigg\langle[\bar{\psi}^{a}(0)
\psi^{b}(0)]
[\bar{\psi}^{c}(x)\gamma^{\mu}\gamma_5
\psi^{d}(x)]
\nonumber\\
&&\hspace{0.5cm}\times[\bar{\psi}^{e}(y)\gamma_{\mu}\gamma_5
\psi^{f}(y)]\bigg\rangle_C\nonumber\\
&&\hspace{0.5cm}=\frac{1}{2}{\cal K}_{11}(\delta^{ad}\delta^{cf}
\delta^{eb}+\delta^{af}\delta^{ed}\delta^{cb})
+{\cal K}_{12}\delta^{ab}\delta^{cf}\delta^{ed}
+\cdots,\nonumber
\end{eqnarray}
\begin{eqnarray}
{\cal K}_{13}&=&\frac{i}{576(N_f^2-1)}\int d x\; \bigg[
(5g_{\mu\nu}g_{\mu'\nu'}-2g_{\mu\mu'}g_{\nu\nu'})
x^{\mu'}x^{\nu'}\nonumber\\
&&\times[\bigg\langle~[\overline{\psi}^a(0)\gamma^{\mu}\psi^b(0)]~
[\bar{\psi}^b(x)\gamma^{\nu}\psi^a(x)]\;\bigg\rangle_C
\nonumber\\
&&
-\frac{1}{N_f}\bigg\langle~[\bar{\psi}^a(0)\gamma^{\mu}\psi^a(0)]
[\overline{\psi}^b(x)\gamma^{\nu}\psi^b(x)]~\bigg\rangle_C]\;\bigg]
\nonumber
\end{eqnarray}
\begin{eqnarray}
&&{\cal K}_{14}
=\frac{i}{36}2g_{\mu\kappa}g_{\nu\lambda}T_A^{\mu\nu\lambda\kappa}
+2g_{\mu\nu}g_{\lambda\kappa}T_A^{\mu\nu\lambda\kappa}
-g_{\mu\lambda}g_{\nu\kappa}T_A^{\mu\nu\lambda\kappa}\nonumber\\
&&\hspace{0.5cm}-2g_{\mu\kappa}g_{\nu\lambda}T_B^{\mu\nu\lambda\kappa}
-2g_{\mu\nu}g_{\lambda\kappa}T_B^{\mu\nu\lambda\kappa}
+g_{\mu\lambda}g_{\nu\kappa}T_B^{\mu\nu\lambda\kappa}\nonumber\label{coeffdef}
\end{eqnarray}
\begin{eqnarray}
{\cal K}_{15}&=&\frac{i}{4(N_f^2-1)}\int d^4 x~ x^{\mu}\bigg[
\bigg\langle \bar{\psi}^a(0)\psi^b(0)
\tan\frac{\vartheta'(0)}{N_f}\nonumber\\
&&\times\bar{\psi}^b(x)\gamma_{\mu}\gamma_5\psi^a(x)\bigg\rangle_C-
\frac{1}{N_f}\bigg\langle\bar{\psi}^a(0)\psi^a(0)
\tan\frac{\vartheta'(0)}{N_f}\nonumber\\
&&\times
\bar{\psi}^b(x)\gamma_{\mu}\gamma_5\psi^b(x)\bigg\rangle_C\bigg]
\label{coeffdef}
\end{eqnarray}
with $T_A,~T_B$ defined as
\begin{eqnarray}                               
&&-\frac{1}{2}\int dxdy\; 
x^{\kappa}\bigg\langle\;
[\overline{\psi}^{a}(0)\gamma^{\mu}\psi^{b}(0)]
[\overline{\psi}^{c}(x)\gamma^{\nu}\gamma_5\psi^{d}(x)]\nonumber\\
&&\times[\overline{\psi}^{e}(y)\gamma^{\lambda}\gamma_5\psi^{f}(y)]
\bigg\rangle_C
\nonumber\\
&&=\delta^{ad}\delta^{cf}\delta^{eb}
T_A^{\mu\nu\lambda\kappa}
+\delta^{af}\delta^{ed}\delta^{cb}
T_B^{\mu\nu\lambda\kappa}+\cdots
\,,\label{TABdef}
\end{eqnarray}
in which the unwritten terms are those 
which do not contribute in (\ref{p4GL}). The integrand in (\ref{p4GL}) is
just the $p^4$-order terms in the Gasser-Leutwyler Lagrangian. 

Eqs.(\ref{F0B0def}), (\ref{F0def}), and  
(\ref{p4C})-(\ref{TABdef}) can be regarded as {\it the fundamental QCD 
definitions of the normal parts of $F^2_0,B_0, L_1,\cdots, L_{10},H_1$ and 
$H_2$}. In principle, the related Green's functions can be calculated from 
lattice QCD. 

As our first calculated result, we present here the values of the
coefficients calculated from solving the equations for the related
Green's functions in a crude approximation. We consider only the
contributions from the two-point Green's functions, and take the
approximations of the large-$N_c$ limit and
the leading order in dynamical 
perturbation theory \cite{PS,HW}. In these approximations, our formula
(\ref{F0def}) for $F^2_0$ reduces to the well-known Pagels-Stokar formula 
\cite{PS}. In the present approximation, the QCD effective coupling constant
is approximately $\alpha_s(p^2)\approx g^2_s/\big(4\pi[1+\Pi_G(p^2)]\big)$
\cite{FR}, where $\Pi_G$ is the gluon self-energy. In the evaluation of the 
Green's functions, we further take the approximation 
$\alpha_s[(p-q)^2]\approx\alpha_s[max(p^2,q^2)]$ to reduce 
the integral equations to differential equations. In the calculation,
we take the modified one-loop formula for $\alpha_s$, i.e. 
$\alpha_s(p^2)=\frac{12\pi}{(33-N_f)\ln(p^2/\Lambda^2_{QCD}+\tau)}$ 
\cite{alpha}.
Since renormalization schemes cannot be distinguished in the one-loop formula
and we take $N_f=3$, the value of $\Lambda_{QCD}$ is not to be compared with 
the experimental value of $\Lambda^{(4)}_{\overline{MS}}$.

In the numerical calculations, we take $F_0=87$ MeV \cite{GL2} as input
by which $\tau$ is related to $\Lambda_{QCD}$. We
present here, in Table 1, the numerical results of the complete coefficients
$L_1,\cdots,L_{10},H_1$ and $H_2$ (the normal part plus the
anomaly contribution \cite{Espriu}) for $\Lambda_{QCD}=600,~700,~800$ MeV
(see Ref.\cite{WKWX2} for details).
\begin{center}
{\small Table 1. Values of the complete $p^4$-order coefficients (in
$10^{-3}$) for various values of $\Lambda_{QCD}$, and the comparison with the 
values determined by experiments from Ref.\cite{GL}.}
\null\vspace{0.2cm}
\doublerulesep 0.6pt
\tabcolsep 0.1in
\begin{tabular}{||c|c|c|c|c||}
\hline\hline
Complete &\multicolumn{3}{|c|}{$\Lambda_{QCD}~({\rm MeV})$}
 &\multicolumn{1}{|c||}{Expt.}\\
\cline{2-4}
Coefficients&600&700&800&\\ \hline
$L_1$ & 1.0 & 1.0 & 1.0 & 0.9$\pm$ 0.3 \\ \hline
$L_2$ & 2.1 & 2.1 & 2.1 & 1.7$\pm$ 0.7 \\ \hline
$L_3$ & -3.1 & -2.5 & -2.0 & -4.4$\pm$ 2.5 \\ \hline
$L_4$ & 0 & 0 & 0 & 0$\pm$ 0.5 \\ \hline
$L_5$ & 4.0 & 4.7 & 5.4 & 2.2$\pm$ 0.5 \\ \hline
$L_6$ & 0 & 0 & 0 & 0$\pm$ 0.3 \\ \hline
$L_7$ & -0.27 & -0.41 & -0.50 & -0.4$\pm$ 0.15 \\ \hline
$L_8$ & 0.006 & 0.44 & 0.69 & 1.1$\pm$ 0.3 \\ \hline
$L_9$ & 6.3 & 6.3 & 6.3 & 7.4$\pm$ 0.7 \\ \hline
$L_{10}$ & -2.4 & -2.1 & -2.1 & -6.0$\pm$ 0.7 \\ \hline
$H_1$ & 1.2 & 1.1 & 1.0 & \\ \hline
$H_2$ & -0.01 & -0.89 & -1.4 & \\ \hline\hline
\end{tabular}
\end{center}
We see from Table 1 that all results are of the right sign and most of them 
are very close to the experimental values except $L_5$, $L_8$ and $L_{10}$. 
Existing certain deviation is understandable since the present approximations 
are rather crude. Improvement of the approach is in progress.

This work is supported by the National Natural Science Foundation of China
and the Fundamental Research Grant of Tsinghua University.
\renewcommand{\baselinestretch}{1}

\end{document}